\documentclass[sigconf]{acmart}

\usepackage{balance}

\usepackage{threeparttable}

\def\BibTeX{{\rm B\kern-.05em{\sc i\kern-.025em b}\kern-.08emT\kern-.1667em\lower.7ex\hbox{E}\kern-.125emX}}
    
\newcommand{\model}{{\em MACDAE}}

\copyrightyear{2019} 
\acmYear{2019} 
\setcopyright{acmcopyright}
\acmConference[KDD'19]{The 25th ACM SIGKDD Conference on Knowledge Discovery and Data Mining}{August 4--8, 2019}{Anchorage, AK, USA}
\acmPrice{15.00}
\acmDOI{10.1145/3292500.3330716}
\acmISBN{978-1-4503-6201-6/19/08}

\begin{document}

\title{Infer Implicit Contexts in Real-time Online-to-Offline Recommendation}

\author{Xichen Ding$^1$, Jie Tang$^2$, Tracy Liu$^2$, Cheng Xu$^1$, Yaping Zhang$^1$, Feng Shi$^1$, Qixia Jiang$^1$, Dan Shen$^1$}
\affiliation{\vspace{0.1cm}
\institution{$^1$ Koubei, Alibaba Group, Beijing, China}
\institution{$^2$ Tsinghua University, Beijing, China}
}
\email{{xichen.dxc, haoze.xc, yaping.zyp, sam.sf, qixia.jqx, dan.sd}@alibaba-inc.com, {jietang,liuxiao}@tsinghua.edu.cn}

\renewcommand{\shortauthors}{Ding et al.}

%

\begin{abstract}

Understanding users' context is essential for successful recommendations, especially for Online-to-Offline (O2O) recommendation, such as Yelp, Groupon, and Koubei\footnote{www.koubei.com, Alibaba's local service company.}. Different from traditional recommendation where individual preference is mostly static, O2O recommendation should be dynamic to capture variation of users' purposes across time and location. However, precisely inferring users' \textit{real-time contexts} information, especially those implicit ones, is extremely difficult, and it is a central challenge for O2O recommendation. In this paper, we propose a new approach, called Mixture Attentional Constrained Denoise AutoEncoder (\model), to infer implicit contexts and consequently, to improve the quality of real-time O2O recommendation. In \model, we first leverage the interaction among users, items, and explicit contexts to infer users' implicit contexts, then combine the learned implicit-context representation into an end-to-end model to make the recommendation. \model\ works quite well in the real system. We conducted both offline and online evaluations of the proposed approach. Experiments on several real-world datasets (Yelp, Dianping, and Koubei) show our approach could achieve significant improvements over state-of-the-arts. Furthermore, online A/B test suggests a 2.9\% increase for click-through rate and 5.6\% improvement for conversion rate in real-world traffic. Our model has been deployed in the product of ``Guess You Like'' recommendation in Koubei.

\end{abstract}

%
%


%

%
\maketitle

\vspace{-8mm}
\subsection*{}
{
\fontsize{8pt}{8pt} \selectfont
\textbf{ACM Reference Format:} \\
Xichen Ding, Jie Tang, Tracy Liu, Cheng Xu, Yaping Zhang, Feng Shi, Qixia Jiang, Dan Shen. 2019. Infer Implicit Contexts in Real-time Online-to-Offline Recommendation. In 
\textit{The 25th ACM SIGKDD Conference on Knowledge} \\
\textit{Discovery and Data Mining (KDD'19), August 4--8, 2019, Anchorage, AK, USA.} ACM, New York, NY, USA, 9 pages. https://doi.org/10.1145/3292500.3330716
}

\section{Introduction}

\begin{figure*}
  \includegraphics[height=3.56in, width=6.6in]{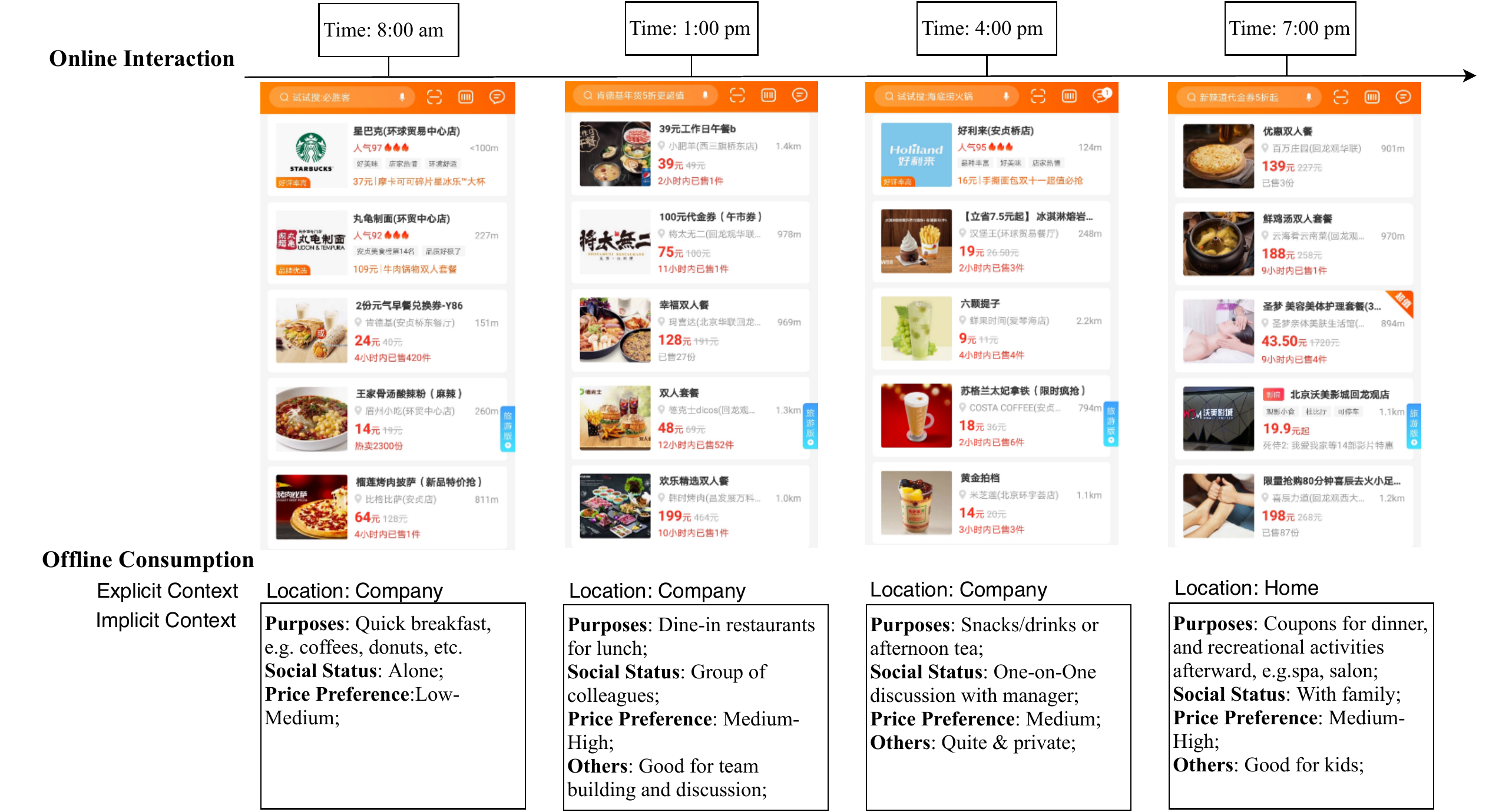}
  \caption{The illustration of implicit contexts in Online-to-Offline recommendation}
  \label{fig:mixture_component}
\end{figure*}

Online-to-Offline (O2O) services, which provide on-demand door-to-door services, have become prevalent in recent years. For example, local business service platforms such as Yelp, Groupon, Dianping, and Koubei, allow users to order products and services online, and receive the products and services offline. Koubei.com belongs to Alibaba's local service company. It serves tens of millions of users everyday by bringing local businesses online and providing various local services to customers, including recommendation of restaurants, local businesses, coupons, etc. 
Users can also place the order through Koubei mobile app in advance before they get to physical place (e.g., the restaurant) and then complete the transaction later. Compared with other e-commence recommendation, O2O  recommendation has several unique characteristics. First, the objective of O2O recommendation is to provide customers real-time recommendation to satisfy their dynamic needs.
This requires us to precisely capture customers'
 dynamic contexts, e.g. their location, time, and status (alone or with friends).
 Though certain contexts such as weekday, time, location, are relatively easy to obtain,
other contexts, which are usually implicit, are  difficult to infer. For example, 
 users' purposes (e.g., whether they are looking for foods or activities),  and status (alone or with friends).
 Figure~\ref{fig:mixture_component} illustrates one example extracted from our data. At a weekday 8:00 am, users may be interested in coupons for breakfast close to their company; around 1:00 pm, they may want to search for deals in a dine-in restaurant; at 4:00 pm, users may want to find a place for afternoon tea. After going back home in the evening, they may switch to search for dinner options and for recreational activities after dinner such as spa and salon. The implicit contexts behind each scenario have multiple dimensions. For example, in the morning the implicit context is to have quick breakfast alone in a cheap restaurant; and for lunch it is to find a place in medium-high price range for group of colleagues. In the afternoon, it changes to find a quite place for discussion,e.g., one-on-one discussion with the manager. In the evening, it is to find restaurants good for family with kids.

Moreover, users' online behaviors at time $t$ may be strongly influenced by other offline contexts. As the example in Figure~\ref{fig:mixture_component} shows, when using Koubei's local business recommendation, users may be interested in items in both current context, representing their real-time dynamic needs, and future contexts, signaling periodical or personal preference. For instance, at 1:00 pm, a customer at work may be interested in both coupons for lunch now, and those for future activities (e.g. dinner for anniversary two days later). In Koubei's ``Guess You Like'' recommendation, our analysis shows that 20.8\% users click on variety of different service providers, including restaurants, gym, spa, etc.

In summary, the fundamental challenge for O2O recommendation, is how to infer users' contexts and how to make accurate recommendations based on the context.
To tackle the problem,  we first infer users' implicit multi-contexts from observational data, including the interactions among users, items, and the explicit contexts. This step is also called pre-training.
Let latent variables \begin{math}[X_{1},X_{2},X_{3},...,X_{n}]\end{math} 
denote one real-time context, where
each variable represents an attribute and their combination describes the context. For example, \begin{math}X_{1}\end{math} denotes users' purposes, \begin{math}X_{2}\end{math} denotes the social status, \begin{math}X_{3}\end{math} denotes price preference, etc.
The implicit multi-contexts can be represented as \begin{math}C_{i}\end{math}.
We propose a novel approach, called Mixture Attentional Constrained Denoise AutoEncoder (\model), to learn the implicit multi-contexts by using generative models.
Assuming that there are \begin{math}K\end{math} multiple implicit contexts, we use \begin{math}C_{ik}\end{math} to denote the \begin{math}k_{th}\end{math} contextual component we want to infer from the observational data. 
We further combine the learned implicit context representation with original input and feed them to downstream supervised learning models, which learn score on the \begin{math}<U, I, C_{e}, C_{i}>\end{math} (User, Item, Explicit Contexts, Implicit Contexts) tuple. 
We compare several generative models in the pre-training stage to infer implicit context, including Denoise AutoEncoder (DAE) \cite{Vincent:2008:ECR:1390156.1390294}, Variational AutoEncoder(VAE) \cite{ICLR:Auto-Encoding_VAE}, and find our proposed model \model\ achieves the best performance for inferring implicit contexts.
We adopt multi-head structure to represent multiple contextual components, in which each head represents one latent contextual component. Furthermore, the importance of each component is learned by attention mechanism. To avoid the problem that different heads learn identical contextual representation, we further apply constraints on the objective function of our generative models.

To summarize, our contributions include: 
\begin{itemize}
\item{\textbf{Implicit Context Modeling}: We take the first attempt to infer users' implicit context from observational data and model implicit multi-contexts in the Online-to-Offline (O2O) recommendation.}


\item \textbf{Context-based Recommendation}: Based on the learned implicit context representations, we present an effective recommendation model using multi-head attentions. 

\item \textbf{High Performance in both Offline and Online Evaluations.}
We conduct both offline and online evaluations to the proposed approach.
Experiments on several real-world datasets show our approach achieves significant improvements over state-of-the-arts. Online A/B test shows 2.9\% lift for click-through rate and 5.6\% lift for conversion rate in real-world traffic. Our model has been deployed in the product of ``Guess You Like'' recommendation in Koubei.


\end{itemize}

The rest of the paper is organized as follows: Section 2 provides an overview of our context-based recommendation system. Section 3 describes details of our proposed model and Section 4 contains results from offline experiments, online A/B test and analysis. We discuss related work in Section 5 and conclude in Section 6. Additional information for reproducibility is provided in supplements.


\begin{figure}
  \includegraphics[height=2.24in, width=3.3in]{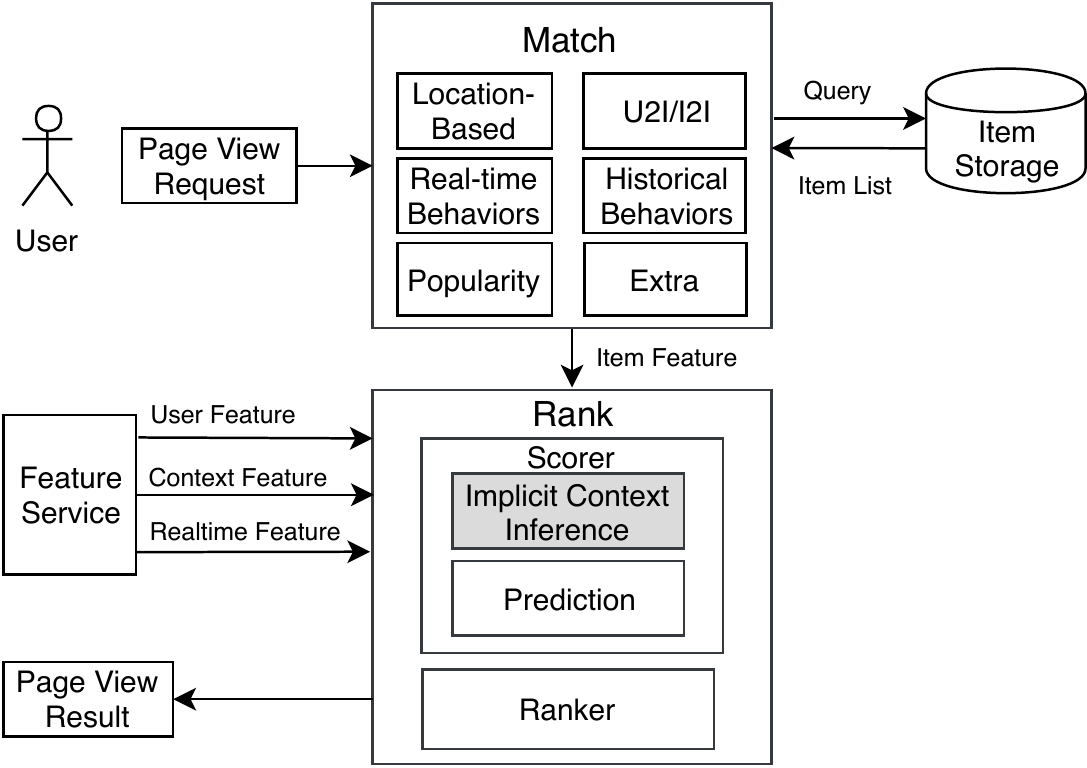}
  \caption{System Overview of Context-Based Recommendation in Koubei}
  \label{fig:system_architecture}
\end{figure}

\section{SYSTEM OVERVIEW}
Figure~\ref{fig:system_architecture} presents the overview of our context-based recommendation system. Users submit a page view (PV) request to the recommendation system. The system first query a set of candidates from storage. There are multiple match strategies, including location-based (retrieve all the items nearby users' current location or objective location), U2I/I2I (user-to-item and item-to-item and other relation based), users' real-time behaviors and historical behaviors, popularity based and others. Second, in the rank step, feature service generates features of users, explicit contexts, and real-time features. These features and item features are fed to the scorer. Simultaneously, users' implicit contexts are inferred from the interaction among users, items and explicit contexts. And in the prediction step, our model predicts a score for each candidate. Finally, candidates are sorted by prediction scores in the ranker. In the following section, we will focus our discussion on the scorer part of the system, especially the implicit context inference step.

\begin{figure*}
  \includegraphics[height=2.4in, width=6.8in]{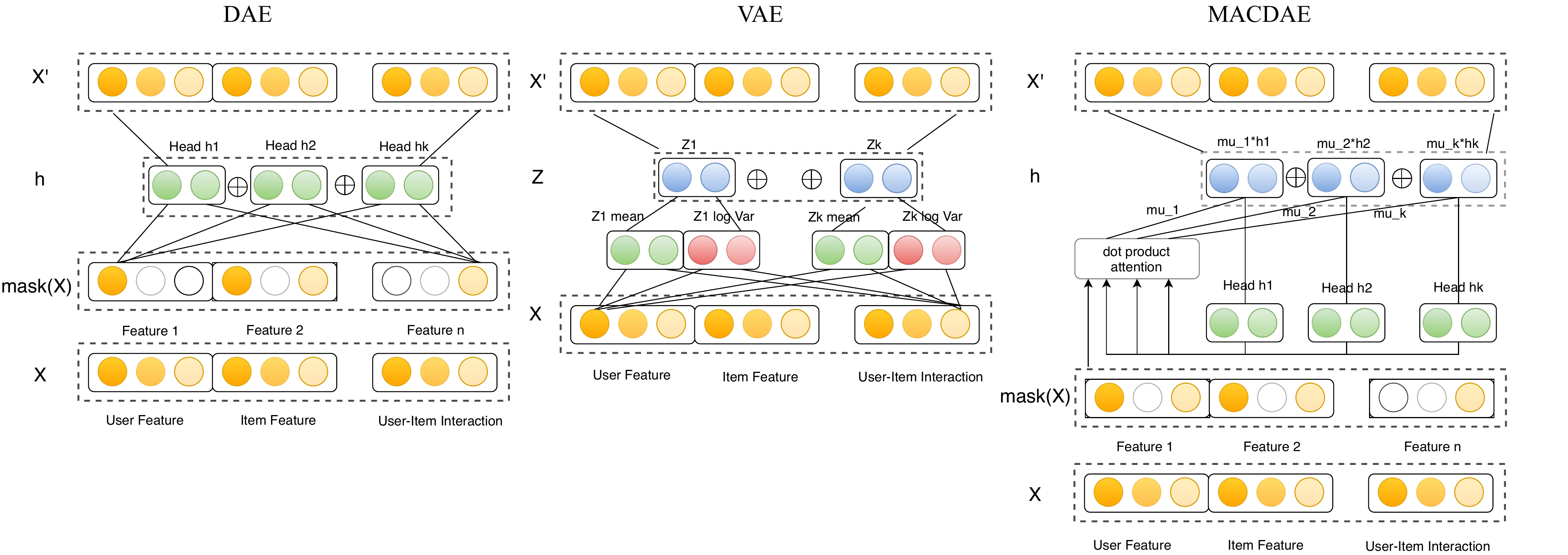}
  \caption{The model architecture of DAE, VAE and MACDAE.}
  \label{fig:model_architect}
\end{figure*}

\section{PROPOSED MODEL}
\subsection{Model Intuition}

There are several approaches to model the implicit multi-contexts in Online-to-Offline recommendation. One straightforward approach is to enumerate all the possible combination of factors that influence users' behaviors, such as weekday, time interval, location, social status and purposes. This approach works when the number of combinations is small, such as finding restaurants for dinner (purpose) in the evening (time interval) near home (location) which is good for family members including kids (social status). However, in practice, there are often too many scenarios that the number of combinations will explode with the growing number of factors.

This leads us to another approach, i.e., inferring users' implicit context as hidden state representation from the observational data, including user, item and explicit context. In addition, the hidden state representation is not a single static one and should reflect the multi-contexts characteristic in the complex Online-to-Offline recommendation. We will discuss several instances of this approach in the model description section.

\subsection{Problem Formulation}
We formulate our implicit multi-contexts modeling problem for recommendation as follows: Let \begin{math}C_{i}\end{math} denote the implicit multi-contexts representation. Basically, we want to infer implicit contexts from the three-way interaction of tuple \begin{math}<U,I,C_{e}>\end{math} (User, Item, Explicit Context). 
Assuming that there are \begin{math}K\end{math} different contextual components in the current implicit context. We use \begin{math}C_{ik}\end{math} to denote the \begin{math}k_{th}\end{math} contextual component we want to infer, which is a \begin{math}d_{k}\end{math} dimension vector. The final implicit contextual representation is:

\begin{equation}
  C_{i}=\sum_{K}^{ }\mu _{k}C_{ik}
\end{equation}
\begin{equation}
  C_{ik}=g_{k}(U,I,C_{e})
\end{equation}
\begin{equation}
  y_{ui}=f(U,I,C_{e},C_{i})
\end{equation}

The \begin{math}\mu_{k}\end{math} denotes the weight importance of the \begin{math}k_{th}\end{math} component, with \begin{math}\sum_{K}^{ }\mu _{k}=1\end{math}. The function \begin{math}g_{k}(.)\end{math} denotes the latent representation function.
And the complete recommendation model predicts a score \begin{math}y_{ui}\end{math} for user and each item in the candidates.

We apply different generative models to instantiate \begin{math}g_{k}(.)\end{math}, including Denoise AutoEncoder (DAE)~\cite{Vincent:2008:ECR:1390156.1390294}, Variational AutoEncoder (VAE)~\cite{ICLR:Auto-Encoding_VAE}, and our proposed model Mixture Attentional Constrained Denoise AutoEncoder (MACDAE). In Section~\ref{sec:exp}, we will compare these models in details.

\subsection{Model Description}
For recommendation with implicit feedback, given a set of users \begin{math}U\end{math} and a set of items \begin{math}I\end{math}, we aim to predict a score \begin{math}y_{ui}\end{math} for each pair of user and item in the list of candidates and then recommend the ranked list of items \begin{math}I_{j}\end{math} to user \begin{math}U_{i}\end{math}. To model the implicit multi-contexts in recommendation, we propose a unified framework, which consists of two steps: First, we infer implicit context representation \begin{math}C_{i}\end{math} from interaction data of \begin{math}<U, I, C_{e}>\end{math} (User, Item, Explicit Context) in the pre-training task. Second, we include the extracted implicit context representation \begin{math}C_{i}\end{math} as additional features and combine with original input and predict the score on the tuple \begin{math}<U, I, C_{e}, C_{i}>\end{math} (User, Item, Explicit Context, Implicit Context).

\textbf{Implicit Context Inference in Pre-training.}
In the pre-training stage, the pre-training dataset comes from positive user-item interaction data, such as clicks and buys. We want to find some latent patterns of implicit context from these observational data. Pre-training method is widely used in many deep-learning tasks such as NLP and Computer Vision. In NLP, language models are pre-trained over large corpus to learn good embedding representation of words and sentences \cite{DBLP:journals/corr/abs-1810-04805}. In computer vision, researchers find it’s beneficial to pre-train image with large dataset and fine-tune the last few layers in application specific tasks \cite{Deng09imagenet:a}. There are two main approaches to apply the pre-trained representation to the following supervised tasks: feature-based approach and fine-tuning approach \cite{DBLP:journals/corr/abs-1810-04805}. Feature-based approach will include the representation as additional features and fine-tuning approach tunes the same model architecture in the downstream supervised learning tasks.

Inspired by previous work, we also use pre-training to learn implicit multi-contexts representation from positive user item interaction data. In our model, the implicit context is inferred from (User, Item and Explicit Context) tuple. We denote \begin{math}x\end{math} as the input to pre-training model architecture, \begin{math}g(x)\end{math} as learned implicit context representation. The input \begin{math}x\end{math} consists of user embedding \begin{math}e_{u}\end{math}, item embedding \begin{math}e_{i}\end{math}, their two-way interaction \begin{math}e_{u}\circ e_{i}\end{math} represented by their element-wise product, and other side information of users and items \begin{math}e_{side}\end{math}. 

\begin{equation}
x=concat(e_{u},e_{i},e_{u}\circ e_{i},e_{side})
\end{equation}

In the downstream supervised learning tasks, we adopt different strategies for different datasets, including feature-based approach and fine-tuning approach. For feature-based approach, we include the pre-trained representation \begin{math}g(x)\end{math} in the deep model architecture, initialize \begin{math}g(x)\end{math} with parameters from pre-training task and freeze them during supervised training. For fine-tuning approach we allow the fine tuning of pre-trained parameters of \begin{math}g(x)\end{math}.

\textbf{Multi-Head DAE/VAE.}
Usually, the implicit contexts have multiple components, which cannot be described by a single static representation. We propose to use a multi-head structure of representation \begin{math}h=g(x)\end{math} to represent the multiple implicit contextual components, where  \begin{math}x\end{math} is the input and \begin{math}g(x)\end{math} is the learning function.
The implicit contextual representation \begin{math}h\end{math} is the combination of multiple contextual components. As illustrated in Figure ~\ref{fig:model_architect}, the hidden layer \begin{math}h\end{math} is the concatenation of \begin{math}K\end{math} multiple heads \begin{math}h_{k}\end{math}, each representing one contextual component \begin{math}C_{ik}\end{math}. Each head (hidden state) \begin{math}h_{k}\end{math} only captures partial contextual information from subspaces of the original input \begin{math}x\end{math}. We denote \begin{math}K\end{math} as the assumed number of contextual components, and use \begin{math}h_{k}\end{math} to represent the \begin{math}k_{th}\end{math} latent hidden contextual component \begin{math}C_{ik}\end{math}. 
For the generative model,  Denoise AutoEncoder(DAE)~\cite{Vincent:2008:ECR:1390156.1390294} and Variational AutoEncoder(VAE)~\cite{ICLR:Auto-Encoding_VAE}, we simply concatenate multiple heads as the final implicit context representation. 

The classic Auto-Encoder\cite{Bengio:2006:GLT:2976456.2976476} model is a one-hidden layer neural network, which first maps an input vector \begin{math}x\end{math} to a latent representation \begin{math}h\end{math} and then reconstruct the input as \begin{math}x^{'}\end{math}. The objective is to minimize the squared loss between the original input \begin{math}x\end{math} and the reconstruction \begin{math}x^{'}\end{math}. The Denoise AutoEncoder\cite{Vincent:2008:ECR:1390156.1390294} further masks out partial information of input with probability \begin{math}p\end{math} and gets the corrupted version of input as \begin{math}\tilde{x}\end{math}. \begin{math}\tilde{x}=mask(x)\end{math}. Then the network aims to encode the corrupted input \begin{math}\tilde{x}\end{math} as hidden state and to reconstruct the original input \begin{math}x\end{math} from the hidden state. As illustrated in Figure~\ref{fig:model_architect}, for the multi-head version of DAE, the hidden state \begin{math}h\end{math} is concatenation of $K$ components. \begin{math}N\end{math} denotes the total number of examples. 

\begin{equation}
h=h_{1} \oplus h_{2} \oplus ...\oplus h_{K},\ k\in K
\end{equation}
\begin{equation}
h_{k}=\sigma (W_{k}\tilde{x}+b_{k}),\ k\in K
\end{equation}
\begin{equation}
x^{'} = \sigma(W^{'}h+b^{'})
\end{equation}
\begin{equation}
L_{reconstruct}=\frac{1}{N}\sum_{N}^{ }||x-x^{'} ||^{2} 
\end{equation}

Variational AutoEncoder (VAE)~\cite{ICLR:Auto-Encoding_VAE} is another very important generative deep learning model, which assumes that the data distribution of input \begin{math}x\end{math} is controlled by a set of latent random variables \begin{math}z\end{math}.

\begin{equation}
p(x)=\int ^{ }p_{\theta }(x|z)p(z)dz
\end{equation}

To optimize \begin{math}p(x)\end{math}, VAE assumes that the latent variable comes from a simple distribution, usually \begin{math}N(0,I)\end{math} and the conditional probability \begin{math}p_{\theta}(x|z)\end{math} is also a Gaussian distribution \begin{math}N(f(z;\theta ),\sigma ^{2}\times I)\end{math} with mean \begin{math}f(z;\theta )\end{math} and covariance \begin{math}\sigma ^{2}\times I\end{math}.  
An encoder (MLP) learns the mean and variance of data distribution and then a decoder reconstructs the input. In the encoder part, since the posterior of \begin{math}p_{\theta}(z|x)\end{math} is intractable, VAE use \begin{math}q_{\phi} (z|x)\end{math} to approximate the original \begin{math}p_{\theta}(z|x)\end{math}. Consequently, the overall structure of VAE becomes an encoder \begin{math}q_{\phi} (z|x)\end{math} and a decoder \begin{math}p_\theta(x|z)\end{math}. The objective function also consists of two parts: the reconstruction part and the Kullback-Leibler divergence \begin{math}D_{KL}\end{math} between \begin{math}q_{\phi} (z|x)\end{math} and \begin{math}p_\theta(z)\end{math}. As shown in Figure~\ref{fig:model_architect}, we also concatenate the state representation of each head \begin{math}z_{k}\end{math} to get the hidden representation \begin{math}z=z_{1} \oplus z_{2} \oplus ...\oplus z_{K}\end{math}, and reconstructs the original input from \begin{math}z\end{math}. For the following supervised training step, we feed the concatenation of mean state vector of each head \begin{math}\bar{z}=\bar{z_{1}} \oplus \bar{z_{2}} \oplus ...\oplus \bar{z_{K}}\end{math} to the prediction model.

\begin{equation}
z=z_{1} \oplus z_{2} \oplus ...\oplus z_{K},\ k\in K
\end{equation}
\begin{equation}
z_{k} \sim  N(\mu_{k}(x),\Sigma _{k}(x)),\ k\in K
\end{equation}
\begin{equation}
x^{'} = \sigma(W^{'}z+b^{'})
\end{equation}

\textbf{Mixture Attentional Constrained Denoise AutoEncoder.}
We propose a new model: Mixture Attentional Constrained Denoise AutoEncoder (MACDAE) (Cf. Figure ~\ref{fig:model_architect}) to infer the multiple implicit contexts. 
Different from the standard Denoise AutoEncoder(DAE) model, in MACDAE, the hidden layer \begin{math}h\end{math} is the concatenation of \begin{math}K\end{math} weighted multiple heads \begin{math}h_{k}\end{math}, with each representing one contextual component \begin{math}C_{ik}\end{math}. 
The 
basic idea is that different implicit contextual components contribute differently to the final representation.
The multi-head DAE model can be considered as a special case of the weighted concatenation where different components contribute equal weights. The weight of each component in MACDAE can be learned using the attention mechanism.

\begin{equation}
h=\mu_{1}h_{1} \oplus \mu_{2}h_{2} \oplus ...\oplus\mu_{K}h_{K},\ k\in K
\end{equation}

The final implicit multi-contexts representation \begin{math}h\end{math} is the concatenation of weighted heads \begin{math}h_{k}\end{math}.  The weight \begin{math}\mu_{k}\end{math} is learned by an attentional function, which maps the uncorrupted version of input \begin{math}x\end{math} to the \begin{math}k_{th}\end{math} hidden representation \begin{math}h_{k}\end{math}. 
In the implementation, 
we use the dot-product (multiplicative) attention. The $(Q,K,V)$ tuple of the attention function is that: the query $Q$ is represented by \begin{math}(W_{a}x)^{T}\end{math} and the keys and values are multiple hidden components \begin{math}h_{k}\end{math}. The original input \begin{math}x\end{math} has the dimension of \begin{math}d_{m}\end{math} and after multiplication with matrix \begin{math}W_{a}\end{math}(shape \begin{math}[d_{k},d_{m}]\end{math}) the dimension of \begin{math}W_{a}x\end{math} becomes \begin{math}d_{k}\end{math}, which has the same dimension as that of the hidden component \begin{math}h_{k}\end{math}. The dimension is also equal to total hidden state dimension \begin{math}d_{h}\end{math} divided by number of heads \begin{math}K\end{math}, \begin{math}d_{k}=d_{h}/K\end{math}. The multiple hidden states \begin{math}h_{k}\end{math} are packed into matrix \begin{math}H\end{math}(shape \begin{math}[K,d_{k}]\end{math}) and the attention function becomes as below. The reconstruction layer is the same as the standard DAE model.

\begin{equation}
Q=(W_{a}x)^{T}
\end{equation}
\begin{equation}
[\mu _{1},\mu _{2},...,\mu _{K}]=softmax(QH^{T}), k\in K
\end{equation}

\textbf{Constraint Definition.}
The loss function of the proposed mixture attentional constrained model uses the squared loss between the original input \begin{math}x\end{math} and the reconstruction \begin{math}\tilde{x}^{'}\end{math}, which is similar to the standard DAE. As we stated previously, one downside of multi-head attention training is the homogeneity between each pair of heads, which suggests that they tend to learn identical representations given the same input \begin{math}x\end{math} and learning steps. Empirically, the average cosine similarity between multiple heads can be as high as 0.788, as illustrated in Figure~\ref{fig:constraint_avg_cosine}, from experiments on the Koubei dataset.

Another idea is to apply constraint on the cosine similarity between each pair of heads of multiple contextual components. Denote \begin{math}h_{i}\end{math} and \begin{math}h_{j}\end{math} as the \begin{math}i_{th}\end{math} and \begin{math}j_{th}\end{math} heads (hidden representation), and the constraint is formulated as:

\begin{equation}
\cos (h_{i},h_{j})\leq \varepsilon, \forall i,j\in K
\end{equation}

\noindent where \begin{math}\varepsilon\end{math} is a hyperparameter, which is within the range \begin{math}\varepsilon<1\end{math}. Since two identical vectors with \begin{math}0\end{math} degree angel has cosine similarity as \begin{math}\cos 0=1\end{math}.  \begin{math}\varepsilon\end{math} denotes the maximum cosine similarity we set on each pair of \begin{math}K\end{math} components, e.g. \begin{math}\varepsilon=0.75\end{math}. We also apply other distance measure, e.g. the euclidean distance between \begin{math}h_{i}\end{math} and \begin{math}h_{j}\end{math}, and find that using cosine distance outperforms the euclidean distance in our experiments.

Now the problem becomes a constrained optimization problem. We want to minimize the reconstruction loss \begin{math}L_{reconstruct}\end{math} subject to the total \begin{math}C_{K}^{2}\end{math} constraints on multiple heads.

\begin{equation}
min\ L=L_{reconstruct}
\end{equation}
\begin{equation}
s.t.\ cos(h_{i},h_{j})-\varepsilon \leq 0,  \forall i,j\in K
\end{equation}

To apply the constraint on the parameters of neural networks, we transform the formulation by adding penalty term to the original objective function and get the final objective function \begin{math}L_{new}\end{math}. \begin{math}\lambda\end{math} is a hyper-parameter of the penalty cost we set on the constraint term of similarity.

\begin{equation}
min\ L_{new}=L_{reconstruct}+\sum_{i,j\in K}{}\lambda (cos(h_{i},h_{j})-\varepsilon )
\end{equation}

\textbf{User Sequential Behavior Modeling.}
For real-world recommendation, there are plenty of user behavior information in logs, such as query, click, purchase, and etc. In Koubei's recommendation, we use a RNN network to model users' sequential behaviors data, with attention mechanism applied to the hidden state of each time step \begin{math}h_{j}\end{math}. We use \begin{math}J\end{math} to denote the total number of steps of users' sequential behaviors. \begin{math}C_{e}\end{math} denotes the explicit contextual information when user's certain behavior actually happens. Let's denote \begin{math}a_{ij}\end{math} as attentional weights that candidate item \begin{math}I_{i}\end{math}, denoted as \begin{math}x_{item_{i}}\end{math}, places on \begin{math}j_{th}\end{math} users' sequential behavior \begin{math}h_{j}\end{math}, which is conditional on the context information \begin{math}C_{e}\end{math}. The final representation of user's sequential behavior given current item \begin{math}I_{i}\end{math}, denoted by \begin{math}x_{attn_{i}}\end{math}, is the weighted sum of the hidden state of RNN as \begin{math}x_{attn_{i}}=\sum_{J}^{ }a_{ij}\times h_{j}\end{math}. The fixed-length encoding vector \begin{math}x_{attn_{i}}\end{math} of users' sequential behavior data under current context \begin{math}C_{e}\end{math} is also combined with the original input \begin{math}x\end{math}, implicit context modeling \begin{math}g(x)\end{math}, and fed to the first layer input of the deep part in supervised learning model \begin{math}x_{0}=concat(x, g(x), x_{attn_{i}})\end{math}.

\begin{equation}
a_{ij}=f_{attn}(x_{item_{i}},h_{j},C_{e})
\end{equation}

\textbf{Learning and Updating of the Two-stage Framework.}
Our proposed framework consists of pre-training stage and downstream supervised learning stage.
We denote \begin{math}N\end{math} as the duration of pre-training dataset, e.g. \begin{math}N\end{math} consecutive days of positive user-item interaction data.  \begin{math}M\end{math} denotes the duration of training dataset, including both impressions and clicks/buys data. The advantage of unsupervised pre-training is that it allows much larger pre-training dataset than the supervised learning dataset, which means \begin{math}N>>M\end{math}. The pre-training model is updated on a weekly basis. It does not need to be updated frequently or by online learning, which saves abundant of computing resources. As for the downstream supervised learning task, since fine-tuning and feature-based approaches are usually faster than learning everything from scratch, we are updating the prediction model once a day after midnight. 

\section{Experiment}
\label{sec:exp}

We conduct both offline and online evaluations to the proposed approach.
In this section, we first introduce the offline evaluation on three different datasets,
and then use the online A/B test to evaluate our proposed model.

\begin{table}
  \setlength{\abovecaptionskip}{0.5pt}
  \caption{Statistics of Yelp, Dianping, Koubei Datasets}
  \label{tab:statistics}
  \begin{tabular}{|c|c|c|c|}
    \hline
    Dataset & User & Item & Interaction \\
    \hline
    Yelp & 24.8K & 130.7K & 1.2M \\
    \hline
    Dianping & 8.8K & 14.1K & 121.5K \\
    \hline
    Koubei* & 4.0M & 1.1M & 43.4M \\   
    \hline
\end{tabular}
\begin{tablenotes}
    \footnotesize
    \item[1] *Koubei offline dataset is randomly sampled from larger production dataset.
\end{tablenotes}
\end{table}

\subsection{Experiments Setup}

\textbf{Dataset.}
We choose three datasets to evaluate our models: Yelp Business dataset\footnote{https://www.kaggle.com/yelp-dataset/yelp-dataset}, Dianping dataset\footnote{http://shichuan.org/HIN\_dataset.html}\cite{10.1007/978-3-319-57529-2_11} and Koubei dataset. Table \ref{tab:statistics} presents the statistics for each dataset. M denotes million and K denotes thousand.

\begin{itemize}

\item{\textbf{Yelp Dataset}
The Yelp dataset (Yelp Dataset Challenge) contains users' review data. It is publicly available and widely used in top-N recommendation evaluation. User profile and business attribute information are also provided. Following \cite{Hu:2018:LMB:3219819.3219965,DBLP:journals/corr/abs-1803-05170}, we convert the explicit review data to implicit feedback data. Reviews with four or five stars  are rated as 1 and the others are rated as 0. Detailed train/test split methods are provided in supplements.
}
\item{\textbf{Dianping Dataset}
Dianping.com is one of the largest  business review websites.  The dataset we use contains customer review data (similar to Yelp Dataset). We also convert review to implicit feedback and prepare the dataset the same way as Yelp dataset. 
}

\item{\textbf{Koubei Dataset}
Koubei.com belongs to the local service company of Alibaba. The goal of the recommendation is to predict user's click-through rate (CTR) and conversion rate (CVR) of candidate items. To evaluate our implicit contextual representation models on this large real-world commercial dataset, we collect 30 consecutive days of positive instance (user click) for pre-training and use 14 days of customer logs data, both positive (user click) and negative (impression without click), for training, and use the data in the next day for testing. For offline evaluation, we randomly sample a subset of a larger production dataset of CTR.}
\end{itemize}

\textbf{Evaluation Metrics.}
For the first two datasets, i.e., Yelp and Dianping, we use Normalized Discounted Cumulative Gain at rank $K$ (\begin{math}NDCG@K\end{math} ) as metrics for model evaluation, which is widely used in top-N recommendation with implicit feedback~\cite{Hu:2018:LMB:3219819.3219965}. For negative sampling of Yelp and Dianping, we randomly sample 50 negative samples for 1 positive sample and then rank the list. The ranked list is evaluated at \begin{math}K=5, 10 \end{math} and the final results are averaged across all users. For the Koubei dataset, we have more information, so we use the Area Under the ROC curve(\begin{math}AUC\end{math}) of the user's CTR as the metric~\cite{DBLP:journals/corr/abs-1803-05170}, which measures the probability that the model ranks randomly sampled positive instance higher than negative instance.

\begin{table*}
  \caption{Results on evaluation of three datasets."*" indicates best performing model and "\#" indicates best baseline model.}
  \label{tab:experiment_result}
  \begin{tabular}{c|c|c|c|c|c}
    \hline
    Model & \multicolumn{2}{|c|}{Yelp} & \multicolumn{2}{|c|}{Dianping} & Koubei \\
    \hline
     & NDCG@5 & NDCG@10 & NDCG@5 & NDCG@10 & AUC \\
    \hline
    Base(Wide\&Deep) & 0.3774 & 0.4272 & 0.4301 & 0.4691 & 0.6754 \\
    \hline
    DeepFM & \#0.3893 & \#0.4414 & \#0.4564 & 0.4854 & 0.6660 \\
    \hline
    NFM & 0.3813 & 0.4242 & 0.4527 & \#0.4996 & \#0.6881 \\
    \hline
    Base+DAE & 0.4159 & 0.4639 & 0.4507 & 0.4954 & 0.6810 \\
    \hline
    Base+VAE & 0.4192 & 0.4661 & 0.4517 & 0.4985 & 0.6790 \\
    \hline
    Base+MACDAE(K=4) & *0.4410 & *0.4886 & 0.4569 & 0.5024 & 0.6902 \\
    \hline
    Base+MACDAE(K=8) & 0.4136 & 0.4650 & *0.4614 & *0.5050 & 0.6936 \\
    \hline
    Base+MACDAE(K=16) & 0.4058 & 0.4521 & 0.4470 & 0.4960 & *0.6954 \\
    \hline
\end{tabular}
\end{table*}

\subsection{Comparison Methods}
We conduct pre-training and compare three models, including DAE, VAE and our proposed \model. 
After we get the representation of implicit context, we adopt different strategies for different datasets. For Koubei dataset, we combine the latent features \begin{math}h=g(x)\end{math} (hidden state) with the original input \begin{math}x\end{math} as additional features, and feed \begin{math}(x+g(x))\end{math} to the baseline model. Then the model is trained with the parameters of \begin{math}g(x)\end{math} frozen. As for Yelp and Dianping dataset, we just use the latent features \begin{math}g(x)\end{math} directly as input and feed to the baseline model.
Since the latent contextual features of Dianping and Yelp are concatenation of embeddings, the final model is fine-tuned and the parameters of the generative model \begin{math}g(x)\end{math} is not frozen. We choose the Wide\&Deep\cite{DBLP:journals/corr/ChengKHSCAACCIA16} model as the baseline model, and compare other models below.

\begin{itemize}
\item{\textbf{Wide\&Deep (BaseModel) }:
The baseline model Wide\&Deep\cite{DBLP:journals/corr/ChengKHSCAACCIA16} combines the wide linear model with deep neural networks. The outputs of the wide part and the deep part are combined using weighted sum and then the model applies a sigmoid function to the final output.
}
\item{\textbf{DeepFM}:
DeepFM\cite{DBLP:journals/corr/GuoTYLH17} model combines two components, Deep Neural Network component and Factorization Machine (FM) component. The FM component models the second-order interaction, compared to the linear part in Wide\&Deep.
}
\item{\textbf{NFM}:
Neural Factorization Machines (NFM)\cite{He:2017:NFM:3077136.3080777} model combines the linearity of Factorization Machine (FM) to model second-order interaction with the non-linearity of neural network to model higher-order feature interactions. A bi-interaction pooling component, which models the interaction among sparse features, is used as hidden layer in the deep structure.
}
\item{\textbf{BaseModel+DAE}:
The pre-training model uses a multi-head version of Denoise AutoEncoder(DAE)\cite{Vincent:2008:ECR:1390156.1390294}. And the hidden state \begin{math}h\end{math} is a concatenation of K different heads (hidden state) as illustrated in Figure~\ref{fig:model_architect}. Finally, we feed the hidden state learned by DAE to the baseline (Wide\&Deep) model.
}
\item{\textbf{BaseModel+VAE}:
The pre-training model uses a multi-head version of Variational AutoEncoder(VAE)\cite{ICLR:Auto-Encoding_VAE} as shown in Figure~\ref{fig:model_architect}. For the downstream training step, we feed the concatenation of mean state vector of each head \begin{math}\bar{z_{k}}\end{math} to the baseline model.
}

\item{\textbf{BaseModel+MACDAE}:
The pre-training model uses our proposed Mixture Attentional Constrained Denoise AutoEncoder(MACDAE) as the objective.  We use the weighted concatenate of multiple heads as hidden state \begin{math}h=\mu_{1}h_{1} \oplus \mu_{2}h_{2} \oplus ...\oplus \mu_{K}h_{K}\end{math} and feed it to the baseline model. For comparison purpose, we implement MACDAE model in three configurations with different multi-head number K, \begin{math}K = [4, 8, 16]\end{math}.
}
\end{itemize}

\subsection{Results from Pre-training Dataset}
\begin{figure*}
\includegraphics[height=3in, width=6in]{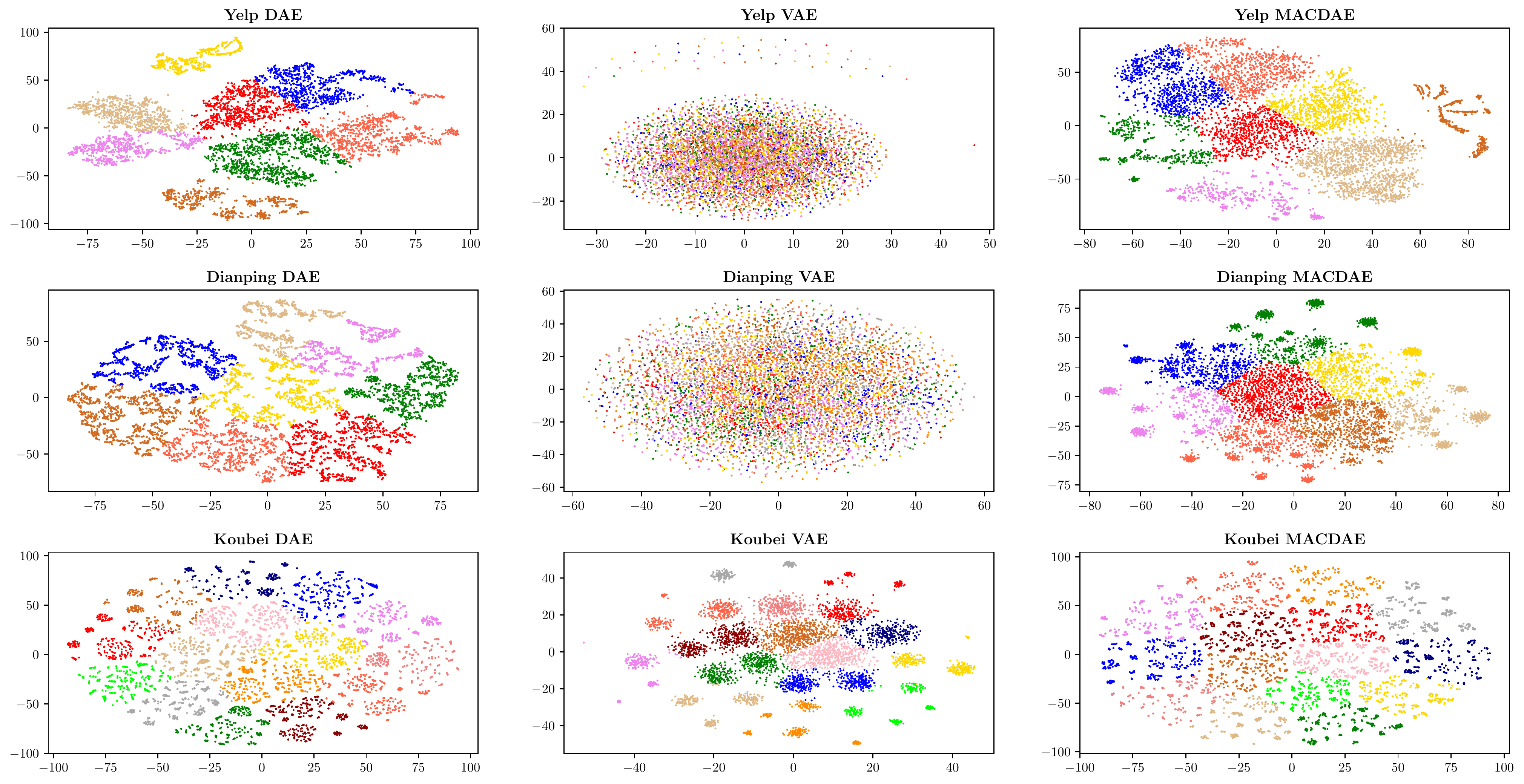}
\caption{The illustration of the latent hidden states of implicit contexts extracted by models}
\label{fig:tsne_latent_space}
\end{figure*}

We have compared DAE, VAE and MACDAE models on three pre-training datasets. The illustration of latent representations learned from the pre-training datasets are presented in Figure~\ref{fig:tsne_latent_space}. The illustration is shown based on 8K samples from Yelp dataset, 8K samples from Dianping dataset and 25.6K samples from Koubei dataset. We use t-SNE\cite{Maaten08visualizingdata} to map the latent representations \begin{math}h\end{math} into 2D space. We also cluster the latent representations and set the total cluster number of Yelp and Dianping to 8 and Koubei to 15. Our illustration shows that the VAE extracts the latent representations as high-variance compared to DAE and MACDAE. The detailed analysis of data distribution will be discussed in the  following section.


\subsection{Offline Model Evaluation Results}
We have conducted extensive experiments on the Yelp, Dianping and Koubei datasets. The Yelp and Dianping datasets are evaluated by NDCG@5 and NDCG@10. The Koubei Dataset of click through rate(CTR) is evaluated by AUC performance. Table~\ref{tab:experiment_result} reports results of model comparison. First, by comparing three fine-tuned baseline models with features from pre-trained models (Base+DAE, Base+VAE, Base+MACDAE) to baseline models without any pretraining, we find that pre-training step and implicit context representations are very effective. For Yelp dataset, compared to Wide\&Deep model without pretraining, the pre-training on DAE, VAE and our proposed MACDAE models gain \begin{math}3.9\%,4.2\%,6.4\%\end{math} absolute improvement on NDCG@5 and \begin{math}3.7\%,3.9\%,6.1\%\end{math} on NDCG@10 respectively. For Koubei CTR dataset, the pre-training step also achieves \begin{math}0.6\%,0.4\%,2.0\%\end{math} absolute improvement on AUC score compared to the baseline model without pretraining. The experiment results are consistent with our hypothesis that adding implicit multi-contexts representations would be helpful for improving the overall performance. Secondly, by comparing different pre-training objectives, our proposed model MACDAE outperforms others, multi-head versions of DAE and VAE. Since multi-head DAE simply uses concatenation of multiple heads, in which all heads contribute equal weights, it is a special case of the general model MACDAE. The attentional weight of multiple components are also effective compared to simple concatenation with equal weights. Moreover, the hyperparameter of multi-head number \begin{math}K\end{math} also influences the performance, and we will discuss the effect of multi-head number \begin{math}K\end{math} in the following section. Thirdly, the results show the effectiveness of adding two-way interaction of user and item features to the input \begin{math}x\end{math} of pre-training model MACDAE. Comparing our proposed model Base+MACDAE with DeepFM/NFM, even if the baseline model (Wide\&Deep) alone performs worse than the DeepFM/NFM due to the loss of high-order feature interaction, our proposed Base+MACDAE still beats these two models. The reason is that the input to the pre-training model MACDAE also contains the features of the two-way interaction between user and item embedding \begin{math}e_{u}\circ e_{i}\end{math}, which is also fed to the baseline model.

\subsection{Online A/B Test Results}


\begin{table}
  \caption{Online A/B Test Results of CTR/CVR in Koubei Recommendation.}
  \label{tab:online_ab_result}
  \begin{tabular}{@{}|c|c|c|}
    \cline{1-3}
    Days & CTR (\%Lift) & CVR (\%Lift) \\
    \hline
    D1 & +3.5\% & +7.0\% \\
    \hline
    D2 & +1.8\% & +6.4\% \\
    \hline
    D3 & +2.6\% & +7.0\% \\
    \hline
    D4 & +2.5\% & +6.3\% \\
    \hline
    D5 & +3.6\% & +2.6\% \\
    \hline
    D6 & +2.9\% & +4.7\% \\
    \hline
    D7 & +3.4\% & +5.3\% \\
    \hline
    Average & +2.9\% & +5.6\% \\
    \hline
\end{tabular}
\end{table}

After comparing the models on Koubei offline dataset, we conduct online A/B test in real-world traffic and compare our proposed model (Wide\&Deep model with MACDAE pre-training) with the baseline model (Wide\&Deep model without pre-training). The daily online results report an average 2.9\% lift on click-through rate(CTR) and 5.6\% lift on conversion-rate(CVR) in Koubei recommendation. The detailed results are reported in table~\ref{tab:online_ab_result}. Online testing indicates that model Base+MACDAE achieves great improvements over the existing online models. More importantly, we have already deployed our new model in production and served tens of millions of customers in Koubei's "Guess You Like" recommendation section everyday.

\subsection{Results from Data Mining Perspective of Context in Recommendation}

\subsubsection{\textbf{Distribution of Latent Representations Learned by Generative Models}}

\begin{figure}
\includegraphics[height=1.7in, width=3.4in]{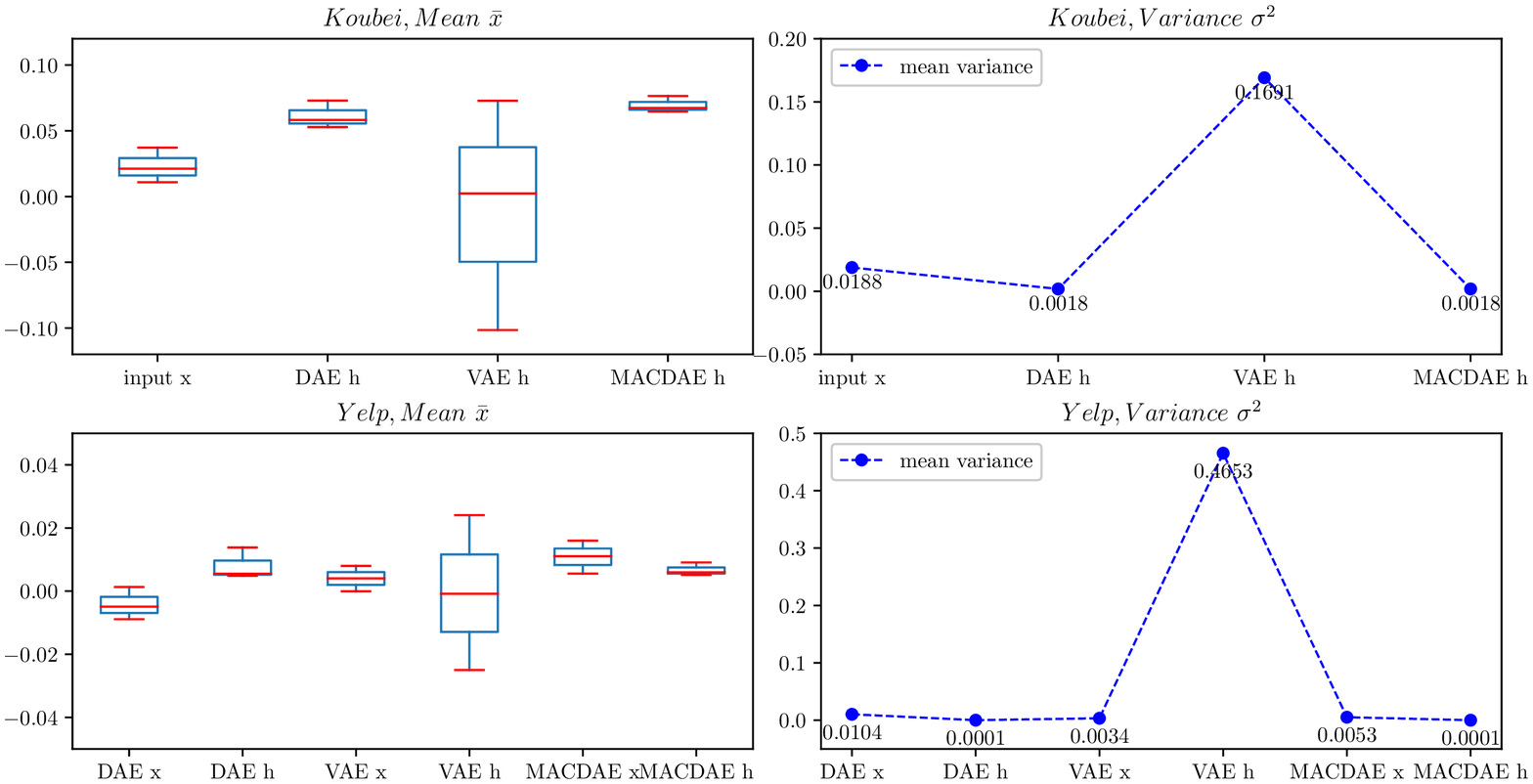}
\caption{Mean and Variance Distribution of Original Input x and Hidden State h Learned by DAE, VAE, MACDAE}
\label{fig:vector_data_dist}
\end{figure}

We conduct experiments to answer the question: What is the data distribution of original input and the latent representations learned by generative models like in recommendation dataset? We compare examples of Koubei dataset with examples of Yelp dataset. Koubei dataset consists of dense features and one-hot encoded sparse features of users and items, and Yelp dataset consists of embedding features learned by user and item id lookup. The results are presented in Figure \ref{fig:vector_data_dist}. We calculate the mean \begin{math}\bar{x}\end{math} and variance \begin{math}\sigma^{2}\end{math} of feature values within each vector of example and the results are then averaged across different samples. We discuss several interesting findings here. First, for Koubei dataset, comparing the same input vector \begin{math}x\end{math} with hidden state \begin{math}h\end{math} learned by different models, DAE model transforms the original input \begin{math}x\end{math} to hidden state \begin{math}h\end{math} by increasing the mean of the features and reducing the variance. In contrast, VAE model transforms the input \begin{math}x\end{math} by reducing the mean and increasing the variance. This is aligned with the illustration of a round-shaped embedding of VAE model trained on Yelp dataset in Figure \ref{fig:tsne_latent_space}. MACDAE is similar to DAE and further reduces the variance in hidden state representation. Second, by comparing VAE model on two different datasets, we find that for Koubei dataset, the hidden state \begin{math}h\end{math} learned by VAE has \begin{math}avg(\bar{h})=0.0023, min(\bar{h})=-0.1015, max(\bar{h})=0.0728, \sigma^{2}=0.1691\end{math}, which shows that the hidden state is not centered around \begin{math}\bar{h}=0.0\end{math}. For the Yelp dataset which is learned from randomly initialized embedding, the hidden state \begin{math}h\end{math} has \begin{math}avg(\bar{h})=-0.0008, min(\bar{h})=-0.0250, max(\bar{h})=0.0241,\sigma^{2}=0.4653\end{math}, which is centered around \begin{math}\bar{h}=0.0\end{math}. Our conjecture is that the original input vector of Koubei dataset consist of one-hot encoded features and dense features, which is very sparse and not normally distributed.

\subsubsection{\textbf{Effect of Multi-Head Number K}}

\begin{figure}
\includegraphics[height=1.7in, width=3.4in]{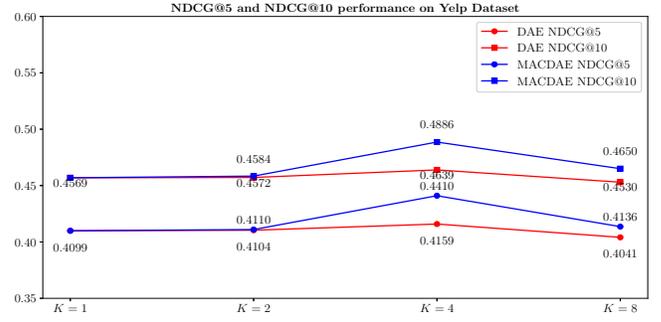}
\caption{The performace of NDCG@5 and NDCG@10 on Yelp dataset with different multi-head number K}
\label{fig:multi_head_k_performance}
\end{figure}

We empirically evaluate the impact of parameter multi-head number K, which represents the assumed hidden state number of the implicit multi-contexts. To highlight the influence of parameter \begin{math}K\end{math}, we choose the Yelp dataset and compare multi-head DAE model with our proposed Mixture Attentional Constrained Denoise AutoEncoder (MACDAE) model on different levels of \begin{math}K\end{math}, \begin{math}K=[1, 2, 4, 8]\end{math}. The performance of evaluation metrics NDCG@5 and NDCG@10 are shown in Figure~\ref{fig:multi_head_k_performance}. The hidden state dimension in Yelp dataset is [256], and  K=4 achieves the best performance of both DAE and MACDAE. For K=1, both DAE and MACDAE become the same vanilla DAE model with single hidden state. Furthermore, increasing head number \begin{math}K\end{math} does not always improve the metrics.

\section{Related Work}

Traditional context-aware recommendation systems use context information as pre-filtering, post-filtering conditions and features in the contextual modeling \cite{Adomavicius:2008:CRS:1454008.1454068}. Methods such as Wide\&Deep\cite{DBLP:journals/corr/ChengKHSCAACCIA16}, DeepFM\cite{DBLP:journals/corr/GuoTYLH17} and NFM \cite{He:2017:NFM:3077136.3080777} only predict score on the \begin{math}<U,I,C_{e}>\end{math} tuple(User, Item, Explicit Context) and neglect the implicit context. Latent factor models including matrix factorization \cite{Koren09matrixfactorization} and tensor factorization \cite{Cunha:2017:MCF:3109859.3109899}, learn hidden factors of user, item, context and their interaction, but only model single latent representation and neglect the characteristic of multi-contexts in O2O recommendation.

Most of recent research on context-aware recommendation systems (CARS) focus on extension of Collaborative Filtering methods~\cite{Zheng12optimalfeature}, Matrix Factorization or Tensor Factorization methods~\cite{Cunha:2017:MCF:3109859.3109899}, and Latent Factor models. For example, HyPLSA\cite{AlanSaid.2009} is high order factorization method that learns interaction among user, item and context. Traditional latent factor models like LDA learns latent subspace from data, but it is time-consuming to train and can't be easily integrated with downstream tasks, such as deep neural networks. Recently pre-training is widely used in NLP, e.g., ELMo\cite{Peters:2018} and BERT\cite{DBLP:journals/corr/abs-1810-04805} models are pre-trained on large corpus and achieve great success in many tasks, e.g., question answering, etc. The building block Transformer \cite{46201} uses multi-head attention to jointly attend to information from different representation subspaces, which has advantages over single head. We adopt similar multi-head structure as Transformer to represent multiple implicit contextual components while Transformer attend on tokens in sentences.

Attentive Contextual Denoising Autoencoder(ACDA) \cite{Jhamb:2018:ACD:3234944.3234956} model extends idea of CDAE~\cite{Wu:2016:CDA:2835776.2835837} and uses a single attentional layer to apply a weighted arithmetic mean on features of the hidden layer representation, which also considers the explicit contextual features. The key difference between our proposed MACDAE and methods mentioned above is that: CDAE and ACDA aim to learn latent features that reflect user's preferences over all N candidate items at the same time and only consider the explicit contextual features. MACDAE infers the implicit contextual representation from the interaction between each pair of user, item and explicit context.

\section{Conclusions}
In this paper, we propose a unified framework to model users' implicit multi-contexts in Online-to-Offline (O2O) recommendation. To infer the implicit contexts from observational data, we compare multiple generative models with our proposed Mixture Attentional Constrained Denoise AutoEncoder (MACDAE) model.
Experiments show that \model\ significantly outperforms several baseline models. Additionally, we conduct extensive analysis on the actual data distribution of latent representations and gain insights on properties of different generative models. Online A/B test  reports great improvements on click-through rate (CTR) and conversion rate (CVR), and we have deployed our model online in Koubei's ``Guess You Like'' recommendation and served tens of millions of users everyday.

\section{Acknowledgements}
We thank algorithm-engineering teams of Koubei and Ant Financial for their support on recommendation platform, feature service, distributed machine learning platform and online model serving system, which enable the successful deployment in production.

%
\bibliographystyle{ACM-Reference-Format}
\bibliography{final_paper_bibliography}

\newpage
\appendix
\section{Supplement}

In this section, we provide details for reproducibility of our experiments and results.

\subsection{\textbf{Implementation Notes}}

\textbf{Dataset Processing}

We have compared multiple generative models on Yelp\footnote{https://www.kaggle.com/yelp-dataset/yelp-dataset}, Dianping\footnote{http://shichuan.org/HIN\_dataset.html}\cite{10.1007/978-3-319-57529-2_11} and Koubei datasets as described in the experiments section. And we want to add more details of the methods we use for dataset processing. 

For Yelp public dataset, we convert the explicit Yelp review data (stars) to implicit feedback data (0/1). We follow  \cite{Hu:2018:LMB:3219819.3219965}\cite{DBLP:journals/corr/abs-1803-05170} to treat reviews with 4 and 5 stars as positive (\begin{math}y=1\end{math}) and the others as negative (\begin{math}y=0\end{math}). We only keep active users with sufficient interactions and filter the subset of users with minimum reviews number as 20. The train/test split is 80\%/20\%. For each user, we randomly sample 80\% of the interactions for training and use the other 20\% for testing. Since evaluating NDCG@K on all candidates is time consuming, we adopt the common negative sampling strategy and set the rate \begin{math}NS=50\end{math}. This means that for each positive user-item interaction data, we randomly sample 50 items which user does not interact with as negative samples. For fair comparison, we prepare the pre-training dataset using only the positive interaction data from the training dataset. 

The Dianping dataset is prepared in the same way as Yelp dataset, except we are using "leave-one-out" strategy for train/test split. For each user, we randomly hold one positive item for testing and the remaining positive items are used for training. Users' minimum reviews number is set to 4 and the negative sampling rate is set to \begin{math}NS=50\end{math}. 

For Koubei dataset, we collect 30 consecutive days of positive instance (user click) for pre-training and use 14 days of user logs data, both positive (user click) and negative (impression without click), for training, and use the data in the next day for testing. For offline evaluation, we randomly sample subset from the production CTR dataset and the statistics of pre-train/train/test instances of the offline dataset is presented in table \ref{tab:statistics_koubei}.

\textbf{Model Configurations and Hyperparameters}

In our experiments, we implement and compare three different generative models during pre-training stage. We implement multi-head DAE and VAE with the same head number \begin{math}K\end{math} as MACDAE for comparison. To compare the effect of different head number \begin{math}K\end{math}, we evaluate the MACDAE model on three different configurations, \begin{math}K = [4, 8, 16]\end{math}. For the Koubei dataset, the dimension of the original input vector \begin{math}x\end{math} is 532, so we set the hidden layer size \begin{math}d_{h}\end{math} to 256 and the number of heads \begin{math}K\end{math} to 8, and each head has dimension of 32. As for Dianping dataset, the dimension of input \begin{math}x\end{math} to encoder is 320, the hidden layer dimension is set to 128 and the number of heads \begin{math}K\end{math} is set to 4 with each head as dimension 32. For Yelp dataset, the input \begin{math}x\end{math} has dimension 403, and we set hidden layer size to 256 with the number of heads \begin{math}K\end{math} to 4. For hyper-parameters of DAE and MACDAE, dropout probability \begin{math}p\end{math} is set to 0.95. For MACDAE, we set the penalty cost \begin{math}\lambda\end{math} to 0.05 for Koubei dataset and 0.005 for Yelp/Dianping dataset. And constraint on cosine similarity \begin{math}\epsilon\end{math} is set to 0.75. We also use Adam as the default optimizer with learning rate set to 0.001. The pre-training for all datasets lasts 5 epochs. And the training for downstream supervised learning model lasts 10 epochs. In the supervised learning task, the Wide\&Deep base model is implemented as follows. For Koubei dataset, the Wide\&Deep model has deep hidden layers as \begin{math}[512, 256, 256]\end{math}. For Yelp dataset, we set the embedding dimension of user and item to 64, deep hidden layers to \begin{math}[256]\end{math}. For Dianping dataset, we also set embedding dimension of user and item to 64 and deep hidden layers to \begin{math}[128]\end{math}.

\textbf{Running Environment}

\begin{table}
  \caption{Statistics of Koubei Dataset}
  \label{tab:statistics_koubei}
  \begin{tabular}{c|c|c|c}
    \hline
    Dataset & Pretrain Instance & Train Instance & Test Instance \\
    \hline
    Koubei & 17.7M & 40.5M & 2.9M \\
    \hline
\end{tabular}
\end{table}

In terms of running environment, the pre-training/training/testing of Yelp and Dianping datasets are performed on the MACBOOK Pro with 2.2 GHz Intel Core i7 8 cores CPU and 16 GB memory, which lasts around 10 hours for pre-training and 9 hours for training. The evaluation of Koubei offline dataset is performed on large-scale distributed systems with 30 workers. For online production model, the pre-training of Koubei production dataset lasts around one day. The training of supervised learning models of Koubei dataset last 9-10 hours everyday. 

\subsection{\textbf{Algorithm and Code Implementation}}

We implement all the algorithms described above using tensorflow\footnote{https://www.tensorflow.org/} in python. And we will describe the code implementation of the models we compared.

\begin{itemize}
\item{\textbf{MACDAE}:
We implement our proposed Mixture Attentional Constrained Denoise AutoEncoder (MACDAE) model by tensorflow API in python. The experimental code is available on GitHub repo(\url{https://github.com/rockingdingo/context\_recommendation}).
}
\item{\textbf{Multi-Head DAE/VAE}:
For baseline comparison models multi-head Denoise AutoEncoder(DAE), and multi-head Variational AutoEncoder(VAE), we slightly modify the original research code of tensorflow (\url{https://github.com/tensorflow/models/tree/master/research/autoencoder}) to the multi-head version.
}
\item{\textbf{Wide\&Deep}:
We follow the official release of Wide\&Deep implementation of tensorflow models (\url{https://github.com/tensorflow/models/tree/master/official/wide\_deep}). Our implementation of  Wide\&Deep model with pre-training slightly modifies the original code. It restores the pre-training model parameters in current session of tensorflow first and the parameters of the prediction model are updated by either feature-based approach or fine-tuning approach.
}
\item{\textbf{DeepFM/NFM}:
For the baseline comparison model, we refer the DeepFM and NFM model implementation in this GitHub repo (
\url{https://github.com/princewen/tensorflow\_practice/tree/master/recommendation}). We make several modifications to the original implementation, such as feature extractor to fit the input of our datasets.
}
\end{itemize}

\subsection{\textbf{Discussions}}

\textbf{Effect of Constraint Threshold and Penalty}

\begin{figure}
\includegraphics[height=1.7in, width=3.4in]{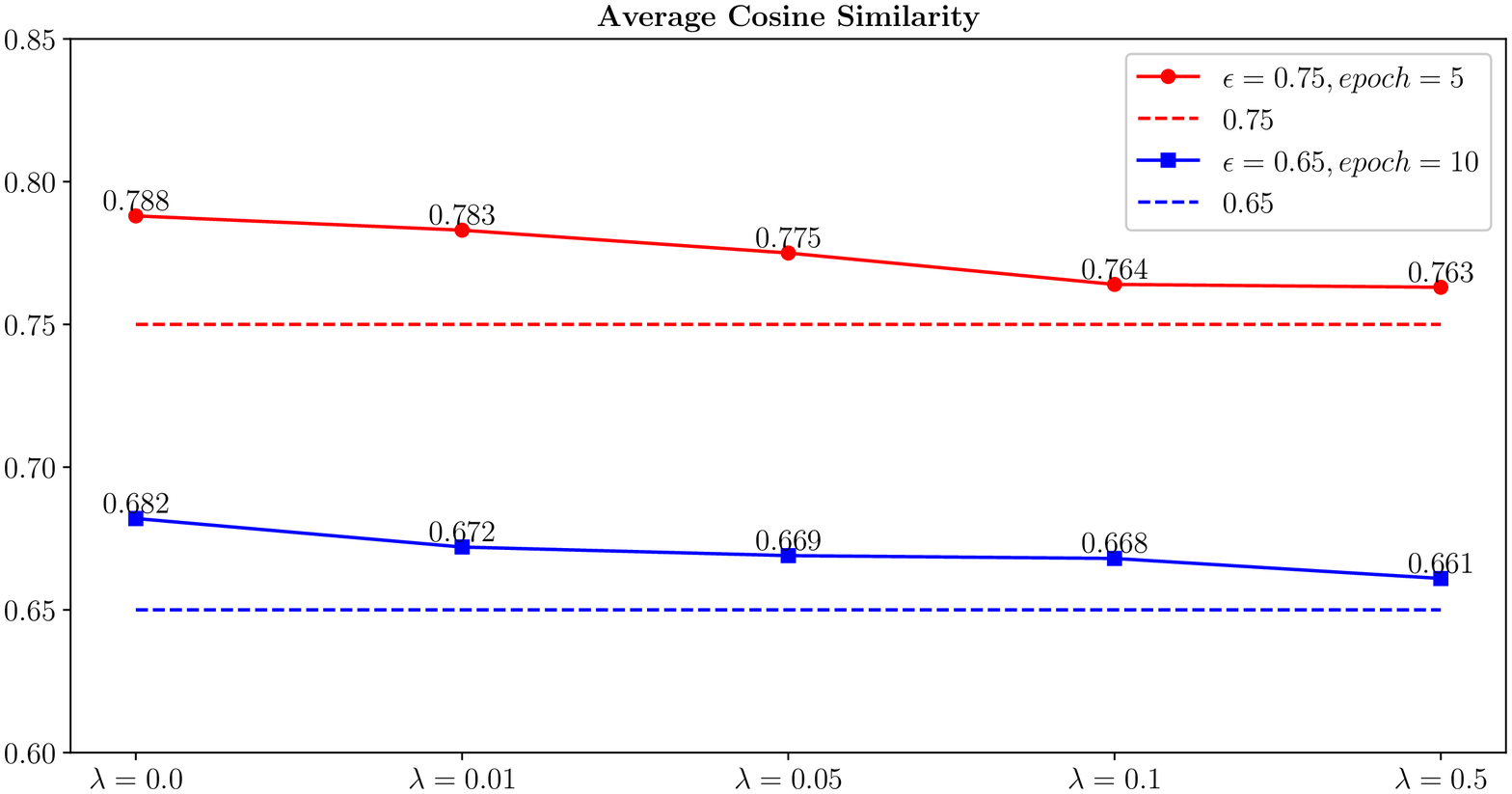}
\caption{Average cosine similarity of multi-heads in MACDAE model pre-trained on Koubei dataset}
\label{fig:constraint_avg_cosine}
\end{figure}

To avoid the problem that multiple contextual components converge to similar latent subspaces, we have applied constraint on the cosine similarity between each heads and add penalty cost to the overall cost function. We set the hyperparameter \begin{math}\epsilon\end{math} as the threshold of maximal cosine similarity between two heads, and \begin{math}\lambda\end{math} as the penalty cost of breaking the constraint in the objective function. Figure~\ref{fig:constraint_avg_cosine} shows the empirical analysis result of the MACDAE model pre-trained on the Koubei dataset with different hyper-parameters. The multi-head number K is set to 8. The penalty cost \begin{math}\lambda\end{math} is set to \begin{math}[0.0, 0.01, 0.05, 0.1, 0.5]\end{math}.  The \begin{math}\epsilon\end{math} is set to 0.75 for 5 pre-training epochs and 0.65 for 10 pre-training epochs. The results show that without applying the constraint, the average cosine similarity among multiple heads can be as high as \begin{math}0.788\end{math}, which means that the angle is close to 0 degree. The average cosine similarity will gradually decrease as we increase the penalty cost \begin{math}\lambda\end{math} and the epoch number.

\textbf{Contextual Feature Importance of Real-world Dataset}

We will present our findings of contextual feature importance of real-world recommendation dataset collected from Koubei's online "Guess You Like" recommendation in Table~\ref{tab:context_feature_importance}. To evaluate the importance of contextual features, we adopt the popular "leave-one-out" strategy, which leaves one contextual feature out and repeat the training process, then evaluate the actual AUC change compared to the baseline model which all the contextual features are available. The dataset is collected on click-through rate(CTR) and conversion-rate(CVR) respectively. CTR measures the percentage of users have impression on the candidates that actually click, and CVR measures the percentage of users click the candidates that actually purchase.

The analyses suggest that time-related contextual features c.time (time intervals of the day, e.g. morning, afternoon, evening, ...), c.weekday (weekdays or weekends) and distance-related features u.s.dist (real-time distance between users' location and recommended shop) rank much higher in AUC contribution on the conversion rate(CVR) dataset than the click-through rate (CTR) dataset. For the click-through rate (CTR) dataset, users' behavioral features, u.s.30d.buy.id (shops that users purchased in the last 30 days) and u.s.30d.clk.id (shops that users clicked in the last 30 days), contribute more than the contextual features and price related features. While for conversion rate (CVR) dataset, the situation is different, where s.price (average transaction price of shop) ranks on the top in AUC contribution. This is aligned with our expectation that users are heavily influenced by their past behaviors of clicks and purchases in the first impression of click-through rate (CTR). Price related features will influence users' decision more in the second conversion stage (CVR), in which clicks convert to actual purchases.

\begin{table}
  \caption{Contextual Feature Importance(AUC change) of CTR and CVR datasets in Decreasing Order}
  \label{tab:context_feature_importance}
  \begin{tabular}{@{}c|@{\;}c@{\;}|c@{\;}|@{\;}c@{}}
    \hline
    \multicolumn{2}{c|}{CTR Dataset} & \multicolumn{2}{|c}{CVR Dataset} \\
    \hline
    Features & AUC Change & Features & AUC Change \\
    \hline
    u.s.30d.buy.id & 0.0458 & s.price & 0.0745 \\
    \hline
    u.age,u.gender & 0.0347 & c.time,c.weekday & 0.0477 \\
    \hline
    u.s.30d.clk.id & 0.0278 & u.s.dist & 0.0463 \\
    \hline
    c.time,c.weekday & 0.0265 & u.s.30d.buy.id & 0.0449 \\
    \hline
    s.price & 0.0263 & u.s.30d.clk.id & 0.0360 \\
    \hline
    u.s.dist & 0.0244 & u.age,u.gender & 0.0231 \\
    \hline
\end{tabular}
\end{table}




\end{document}